\documentclass{JHEP3}
\usepackage{amssymb}
\usepackage{epsfig}
\usepackage{graphicx}

\def\Schw{Schwarzschild}

\newcommand{\be}{\begin{equation}}
\newcommand{\ee}{\end{equation}}
\newcommand{\bea}{\begin{eqnarray}}
\newcommand{\eea}{\end{eqnarray}}

\def\Schw{Schwarzschild }
\def\({\left(} \def\){\right)}

\title{\center{Fundamental Strings and Higher Derivative Corrections to $d$-Dimensional Black Holes}}

\author{Amit Giveon \\
Racah Institute of Physics, The Hebrew University\\
Jerusalem 91904, Israel\\
E-mail: \email{giveon@phys.huji.ac.il}}

\author{Dan Gorbonos \\ Department of Physics, University of Alberta,
\\ Edmonton, Alberta, Canada T6G 2G7 \\
    E-mail: \email{gorbonos@phys.ualberta.ca}}

\author{Merav Stern \\
Racah Institute of Physics, The Hebrew University\\
Jerusalem 91904, Israel\\
E-mail: \email{merav@phys.huji.ac.il}}

\abstract{We study aspects of $d$-dimensional black holes with two
electric charges, corresponding to fundamental strings with generic
momentum and winding on an internal circle. The perturbative
$\alpha'$ corrections to such black holes and their gravitational
thermodynamics are obtained. The latter are derived
using the Euclidean approach and the Wald formula for the entropy.
We find that the entropy and the charge/mass
ratio of black holes increase in $\alpha'$ for any mass and charges,
and in all dimensions.}

\begin{document}

\section{Introduction}

In a theory of quantum gravity it is important to investigate, in
particular, the properties of black holes. Recently, on general
grounds, it was conjectured that a consistent theory of quantum
gravity implies certain corrections to some thermodynamical
properties of black holes \cite{harvard,Kats:2006xp}.
Since string theory is a
candidate for a consistent theory of quantum gravity, in this work
we shall investigate stringy corrections to black holes.

Concretely, we will inspect black holes formed by highly excited
fundamental strings with generic charges. We shall find the leading
order corrections to the geometry of $d$-dimensional black holes in
the inverse string tension, $\alpha'$, in any dimension $d\geq 4$.
Consequently, we shall compute the linear corrections to the
thermodynamical properties of such black holes.

Our main results are that both {\it the entropy and the charge/mass
ratio of black holes increase in $\alpha'$ for any mass and charges,
and in all dimensions}.
The results are consistent with the conjectures of \cite{harvard,Kats:2006xp}.

To derive these results, we begin in section
2 by obtaining the leading order black hole solutions in any
dimension $d\geq 4$, formed by an excited fundamental string with
any momentum $n$ and winding $w$ on an internal circle, and inspect
their thermodynamical properties.
We choose to parameterize these charges in terms of the left
and right handed momenta:
\bea p_{L}=\frac{n}{R}-\frac{w\,R}{\alpha'}~,\nonumber\\
p_{R}=\frac{n}{R}+\frac{w\,R}{\alpha'}~.\label{plpr}\eea
In the following sections we present higher derivative corrections
to these black holes. The $d=4$ cases were studied in~\cite{first}.

In section 3, we review the leading $\alpha'$ corrections
to the Schwarzschild black hole in $d\geq 4$ --
the Callan-Myers-Perry solution~\cite{CMP}.
In section 4, we obtain the $\alpha'$ corrections to
black holes with momentum charge (the $p_L=p_R$ case),
and in section 5 we study the $\alpha'$ corrections for
generic $(p_L,p_R)$.
In section 6, we rederive the entropy by using the Wald formula,
and recapitulate with a few comments regarding the interpretation of
the results in terms of an effective gravitational coupling.
Finally, in two appendices we present some useful expressions and calculations.

\section{$d$-dimensional black holes with $(p_L,p_R)$ charges --
leading order in $\alpha'$} \label{first section} In this section we
generalize to $d>4$ dimensions the solution of a black hole with momentum and
winding charges (two $U(1)$ charges). The solution in the $d=4$ case was
given for example in~\cite{HorPol}.
The gravitational thermodynamics of this solution was also inspected
in~\cite{HorPol}, and we generalize it here to $d>4$ dimensions.

This solution is constructed in a procedure of adding a momentum and
winding charges by ``lifting'' (or sometimes ``oxidating") the
metric to an additional dimension whose coordinate will be denoted
by $x$. Thus we produce a uniform black string. We take the additional
dimension to be compact. In order to add the momentum charge we
perform a boost in the $x$ direction, which after
Kaluza-Klein (KK) reduction gives one $U(1)$ charge. Applying a T-duality
transformation in the $x$ direction gives a ($d+1$)-dimensional black
string winding around the $x$ circle. Then reducing to $d$ dimensions
one obtains a black hole with winding charge. So in order to add
both charges to the Schwarzschild solution one has to add, after the
boost and the T-duality, a second boost of the black string in the $x$
direction. Then a reduction to $d$ dimensions generates the black
hole with generic momentum and winding charges $(p_L, p_R)$~(\ref{plpr}).

The boost and the T-duality are part of the $O(2,2)$ symmetry group
of the low energy effective action
\be
I_{eff}=\frac{1}{16\,\pi\,G_d}\int\!\!d^{d}x\,\sqrt{-g}\,e^{-2\,\phi}\(R+4\,\(\nabla\phi\)^2-\frac{1}{12}H^{2}\)~,
\ee
where
\be
H_{\alpha\beta\gamma}=3\,\partial_{[\alpha}B_{\beta\gamma]~,}
\ee
and $B_{\mu\nu}$ is the Kalb-Ramond field. Performing the above
symmetry transformations map solutions to solutions and in
particular change the value of the scalar field $\phi$ -- the dilaton --
and $B_{\mu\nu}$.
In particular, when we start from a Schwarzschild solution,
the above transformations will turn on the additional fields.

Explicitly,
we start the procedure with the $d$-dimensional Schwarzschild-Tangherlini
black hole metric~\cite{tangher,emparan}
\be ds^{2}=-f(\rho)\,dt^{2}+\frac{1}{f(\rho)}\,d\rho^{2}+\rho^{2}\,d
\Omega_{d-2}^{2}\,,\ee
where
\be f(\rho)=1-\frac{\rho_{s}^{d-3}}{\rho^{d-3}}~,\label{fffff}\ee
$\rho_s$ is related
to the black hole mass $M$ via~\cite{emparan}
\begin{equation}
\rho_{s}^{d-3} = \frac{16\,\pi\, G_{d}\,M}{(d-2)\,\Omega_{d-2}} \,,
\label{rho-m}
\end{equation}
$\,d \Omega_{d-2}^{2}$ is the metric on a unit $S^{d-2}$ and
\be \Omega_{d-2}=\frac{2\,\pi^{\frac{d-1}{2}}}{\Gamma(\frac{d-1}{2})}\ee
is its area.

Next we summarize the transformations that will be used to
generate of the doubly-charged solution. Let us denote by
$g_{\mu\nu}^{\alpha}$ a boosted metric in the additional direction $x$
with a boost parameter $\alpha$. We express $g_{\mu\nu}^{\alpha}$ in
terms of the original metric $g_{\mu\nu}$ as
\bea g_{tt}^{\alpha}&=&\cosh^{2}(\alpha) g_{tt}+\sinh^{2}(\alpha),\nonumber\\
g_{xt}^{\alpha}&=&\sinh(\alpha)\cosh(\alpha)\(g_{tt}+1\), \label{boosted}\\
g_{xx}^{\alpha}&=&\sinh^{2}(\alpha)
g_{tt}+\cosh^{2}(\alpha),\nonumber
 \eea
 where the rest of the metric components and other fields do not change.

A T-duality with respect to the additional direction is given by the
following set of rules (see e.g. \cite{GPR} for a review):
\bea g_{\mu\nu}^{T}&=&g_{\mu\nu}-\frac{g_{\mu x}\, g_{\nu x}-B_{\mu
x}\,B_{\nu x}}{g_{xx}}, \quad \quad g_{\mu x}^{T}=\frac{B_{\mu
x}}{g_{xx}}, \quad \quad g_{xx}^{T}=\frac{1}{g_{xx}}, \nonumber\\
B_{\mu\nu}^{T}&=&B_{\mu\nu}-\frac{B_{\mu x}\,g_{\nu x}-g_{\mu
x}B_{\nu x}}{g_{xx}}, \quad \quad B_{\mu x}^{T}= \frac{g_{\mu
x}}{g_{xx}},\quad\quad
\phi^T=\phi+\frac{1}{2}\,\ln\sqrt\frac{g^T}g, \label{Trules}\eea
where the metric and other field components after the transformation
are denoted by the superscript $T$.

As described above, we take the $d$-dimensional
Schwarzschild-Tangherlini solution, perform a boost with the parameter
$\alpha_{w}$ along the additional compact direction $x$ and then apply
the T-duality rules. After the application of a second boost with
the parameter $\alpha_{n}$ we reduce the solution to $d$ dimensions
and obtain the following $d$-dimensional black hole with fundamental
string charges $(p_L,p_R)$ (in the string frame):
\be
ds^2 = -\frac{f(\rho)}{\Delta\left(\alpha_n\right)\Delta\left(\alpha_w\right)} dt^2 + f(\rho)^{-1} d\rho^2 + \rho^2 d\Omega^2_{d-2},\\
\label{hor-pol_metric} \ee where \be \label{delta}
\Delta\left(x\right)\equiv1+\left(\frac{\rho_s}{\rho}\right)^{d-3}\sinh^2x,
\ee with a dilaton \be \phi(\rho)=\phi_0- \frac{1}{4} \log
\Delta\left(\alpha_n\right) - \frac{1}{4} \log
\Delta\left(\alpha_w\right), \ee where $\phi_0$ is a constant, and
two Abelian gauge fields \be \emph{A}^n_{t} =
\frac{1}{2}\left(\frac{\rho_s}{\rho}\right)^{d-3}\frac{\sinh2\alpha_n}
{\Delta\left(\alpha_n\right)}, \quad \quad \emph{A}^w_{t} =
\frac{1}{2}\left(\frac{\rho_s}{\rho}\right)^{d-3}\frac{\sinh2\alpha_w}
{\Delta\left(\alpha_w\right)}. \label{vec_potent}\ee
The boost parameters $\alpha_{w}$ and $\alpha_{n}$
correspond to the two boosts described above, generating the winding
and momentum charges, respectively. $\emph{A}^{n}$ is the vector
potential which is coupled to the momentum of the fundamental string
and $\emph{A}^{w}$ is coupled to its winding. Note that the horizon of
the black hole is located at $\rho_s$ for any value of $\alpha_{n,w}$.

The conserved charges associated with the vector potentials above
are  \be Q=\frac{1}{16\pi G_d} \int_{S_{\infty}^{d-2}}{\ast\, dA}~,
\ee where $S_{\infty}^{d-2}$ is the ($d-2$)-dimensional sphere at
infinity and $*dA$ is the Hodge dual of the 2-form field strength.
Then
\be Q_n=\frac{\left(d-3\right)\Omega_{d-2}\rho_s^{d-3}}{32\pi G_d}
\sinh2\alpha_n,
 \quad \quad
Q_w=\frac{\left(d-3\right)\Omega_{d-2}\rho_s^{d-3}}{32\pi G_d}
\sinh2\alpha_w. \label{charges}\ee
The left and right moving momenta (\ref{plpr}) are accordingly: \be
p_{L}=Q_{n}-Q_{w}, \quad \quad p_{R}=Q_{n}+Q_{w}. \label{pp} \ee The
corresponding chemical potentials are\bea
\Phi_{L}=\frac{1}{2}\(\tanh(\alpha_{n})-\tanh(\alpha_{w})\),
\label{phi} \quad \quad
\Phi_{R}=\frac{1}{2}\(\tanh(\alpha_{n})+\tanh(\alpha_{w})\),
 \eea
which are equal to the values of electromagnetic potentials
$A_t^{L,R}$ at the horizon, where
$(A_L,A_R)\equiv\frac{1}{2}(A^n-A^w,A^n+A^w)$.

The surface gravity is \be
\kappa=\left.\frac{1}{2}\,\frac{\left|\partial_{\rho}g_{tt}\right|}
{\sqrt{-g_{tt}\,g_{\rho\rho}}}\right|_{\rho=\rho_{s}}
=\frac{d-3}{2\,\rho_{s}\,\cosh\alpha_n\,\cosh\alpha_w}~,
\ee
and then the inverse temperature is \be
\beta=\frac{2\,\pi}{\kappa}=\frac{4\,\pi\,\rho_{s}\,\cosh\alpha_n\,\cosh\alpha_w}{d-3}
\;\;.\label{inv_temp} \ee
We can calculate the Euclidean action for this solution \be
I_E=\beta\,F,\ee where $F$ is the free energy. The integrand of the
Euclidean action is invariant under boost and T-duality. The only
change is in the limits of integration in the plane of the Euclidean
time and the additional compactified dimension. Since the metric is
static, the integration over the Euclidean time gives only a
multiplicative factor by its period $\beta$, and the change in the
size of the compactified direction is absorbed after the KK
reduction by the $d$-dimensional Newton constant. Hence, the boost
and the T-duality do not change the value of the free energy, so we
can take its value for the Schwarzschild-Tangherlini case: \be
 F=\frac{\Omega_{d-2} \rho_s^{d-3}}{16\pi G_d}~,
\ee and then the Euclidean action is \be
I_E=\frac{\rho_{s}^{d-2}\,\Omega_{d-2}\,\cosh\alpha_n\,\cosh\alpha_w}{4\,G_{d}\,(d-3)}.
\ee
 The ADM mass is
\be M= \frac{\partial\( \beta\,F\)}{\partial \,
\beta}+p_{L}\,\Phi_{L}+p_{R}\,\Phi_{R}=
\frac{(d-2)\,\Omega_{d-2}\,\rho_{s}^{d-3}}{16\,\pi\,G_{d}}
\(1+\frac{d-3}{d-2}\(\sinh^{2}(\alpha_n)+\sinh^{2}(\alpha_w)\)\)~,
\label{mass0} \ee
and the entropy is
\be
 S=\beta\,\(M-F-p_{L}\,\Phi_{L}-p_{R}\,\Phi_{R}\)=
\frac{\rho_{s}^{d-2}\Omega_{d-2}\cosh\alpha_n\cosh\alpha_w}{4G_d}.
\label{smplpr} \ee
This expression can be interpreted as the \Schw entropy, which is
proportional to the area of the black string, boosted twice along
the additional direction of the compactified dimension. The inverse
temperature $\beta=1/T$ can be obtained also from
 \be \beta=\(\frac{\partial\,S}
 {\partial\,M}\)_{p_{L},p_{R}}~.
 \label{invttt}\ee
The extremal limit is obtained by taking, say, $p_{R} \rightarrow
M$. This amounts to taking either $\alpha_n\to\infty$ and/or
$\alpha_w\to\infty$, as well as $\rho_s\to 0$, such that
$\rho_{s}^{d-2}(\exp(2\alpha_n)+\exp(2\alpha_w))$ is held fixed. In
this limit the horizon is singular and, in particular, $S \rightarrow 0$.

Let us consider the case when there is no winding charge,
$\alpha_w=0$ in (\ref{hor-pol_metric}). Then the charges, that
correspond to the right and left moving modes, are equal $ p \equiv
p_{L}=p_{R}$. In this case it is possible to write explicitly the
entropy and the temperature as a function of the mass and the
charges by reversing the relations ($\ref{charges}, \ref{mass0}$).
We obtain the following expressions for $\alpha_n$ and $\rho_s$ as a
function of the charge to mass ratio
\be q \equiv \frac{p}{M},\ee
\bea \cosh^2(\alpha_n)&=&\frac{d-3+2\,q^{2}+\delta}{2(d-3)(1-q^{2})}\label{seder01},\\
\rho_s^{d-3}&=&\frac{32\,\pi\,G_d\,M\,(1-q^{2})}{\Omega_{d-2}\,(d-1+\delta)}\label{seder02},
 \eea
where \be \delta \equiv \sqrt{(d-3)^{2}+4\,(d-2)\,q^2}\label{sdelta}.\ee Substituting
into the expression for the entropy ($\ref{smplpr}$) we obtain \be
S=\frac{2^\frac{5d-7}{2(d-3)}\sqrt{\pi}\left[G_{d}\Gamma\(\frac{d-1}{2}\)\right]^{\frac{1}{d-3}}\,M^{\frac{d-2}{d-3}}\sqrt{2\,q^{2}+d-3+\delta}\(1-q^2\)^{\frac{d-1}{2\,(d-3)}}}{\sqrt{d-3}\(d-1+\delta\)^{\frac{d-2}{d-3}}},
\label{entroppy_base}\ee
and substitution into the temperature ($\ref{inv_temp}$) gives \be
T=\beta^{-1}=\frac{(d-3)^{\frac{3}{2}}\(d-1+\delta\)^{\frac{1}{d-3}}\(1-q^{2}\)^{\frac{d-5}{2\,(d-3)}}}{2^{\frac{3d-1}{2(d-3)}}\,\sqrt{\pi}\left[\Gamma\(\frac{d-1}{2}\)G_{d}M\right]^{
\frac{1}{d-3}}\sqrt{2q^{2}+d-3+\delta}} \label{temperature_base}.\ee

\section{The $\alpha'$ corrections to Schwarzschild -- the Callan-Myers-Perry solution}
\label{CMP} We now review the higher derivative gravity corrections
to the Schwarzschild black hole~\cite{CMP}. The leading order correction to the
low energy string effective action $I_{eff}^0$ in $d$ dimensions
gives rise to~\cite{Tseytlin1}
\be I_{eff}^{\lambda}= \frac{1}{16\,\pi\,G_{d}} \int
d^{d}x \,\sqrt{-g}\,e^{-2\,\phi}\,\left(R+4\,(\nabla\phi)^{2}
+\frac{\lambda}{2}\,R_{\mu\nu\rho\sigma}\,R^{\mu\nu\rho\sigma}\right),
\label{Seff}\ee where
$\lambda=\frac{\alpha'}{2},\,\frac{\alpha'}{4},\,0$ for bosonic,
heterotic and type II strings, respectively, and $G_{d}$ is the
Newton constant in $d$ dimensions. Any other form of the correction
to the action at the same order in $\alpha'$ is equivalent to this
one by field redefinitions~\cite{Tseytlin1}.

We shall concentrate on spherical symmetric solutions in $d\geq4$
dimensions. Hence, we take the following ansatz: \be ds^{2}=g_{tt}\,
dt^{2}+g_{\rho\rho}\,d\rho^{2}+\rho^{2}\,d\Omega_{d-2}.
\label{ansatz} \ee The solution to the equations of motion with the
appropriate boundary conditions (regularity at the horizon and
asymptotic flatness as $\rho \rightarrow \infty$) to first order in
$\lambda$ is~\cite{CMP}: \bea g_{tt}&=&-f\,\left(1+2\,\lambda
\,\mu(\rho)\right), \nonumber          \\
g_{\rho\rho}&=&f^{-1}\,\left( 1+2\,\lambda\,\epsilon(\rho)\right),
\label{Scwa corrected}\\
\phi&=&\phi_{0}+\lambda \,\varphi(\rho),\nonumber  \eea where
$\phi_{0}$ is the constant value of the dilaton in the zeroth order
solution in $\lambda$.
{}For $d=4$: \bea
\mu(\rho)&=&-\left(\frac{23}{12\rho_s\,\rho}+\frac{11}{12\,\rho^{2}}
+\frac{\rho_s}{2\,\rho^{3}}\right),\label{mmm}\\
\epsilon(\rho)&=&-\left(\frac{5\,\rho_{s}}{6\,\rho^{3}}
+\frac{7}{12\,\rho^{2}}+\frac{1}{12\,\rho_{s}\,\rho}\right)\label{eee},\\
\varphi(\rho)&=&-\left(\frac{\rho_s}{3\,\rho^{3}}
+\frac{1}{2\,\rho^{2}}+\frac{1}{\rho_{s}\,\rho}\right). \eea For
$d=5$: \bea
\mu(\rho)&=&-\(\frac{9\,\rho_{s}^2}{4\,\rho^4}+\frac{17}{4\,\rho^2}\),\label{mmm5}\\
\epsilon(\rho)&=&-\frac{\rho^2+6\,\rho_{s}^2}{4\,\rho^{4}}\label{eee5}\,,\\
\varphi(\rho)&=&-\frac{9}{8}\,\(\frac{2\,\rho^2+\rho_{s}^2}{\rho^4}\),
\eea
and for $d > 5$: \bea
\varphi(\rho)&=&-\frac{\(d-2\)^2}{4\,\rho_{s}^2}
\(\frac{d-3}{d-1}\,\frac{\rho_s^{d-1}}{\rho^{d-1}}
+\frac{(d-3)\,\rho_s^2}{2\,\rho^2}-K_d\(\frac{\rho}{\rho_s}\)
-u\left(\frac{\rho}{\rho_s}\right)\)~, \label{varphisolution}\\
\epsilon(\rho)&=&\frac{(d-3)\rho_s^{d-5}}{4(\rho^{d-3}-\rho_s^{d-3})}\,\left(\frac{2\,c_d}{d-3}
+\frac{2(2d-3)}{d-1}\,\frac{\rho_s^{d-1}}{\rho^{d-1}}-\frac{(d-2)(d-3)\,\rho_s^{2}}{2\rho^{2}}
\right. \nonumber \\
&+& \left.
(d-2)\left[K_d\(\frac{\rho}{\rho_s}\)+u\(\frac{\rho}{\rho_s}\)\right]\right)~,
\label{endphieps}\eea where $u(x)\equiv
\ln\(1+\frac{1}{x}+...+\frac{1}{x^{d-4}}\)$, and
the function $K_d(x)$ and the constant $c_d$ are given in appendix A.
$\mu(\rho)$ can be obtained from the relation: \be
\mu(\rho)=-\epsilon(\rho)+\frac{2}{d-2}\(\varphi(\rho)-\rho\,\varphi'(\rho)\)~.
\label{endmu}\ee

We can write the periodicity $\beta$ of the Euclidean time using the
surface gravity at the horizon \be
\kappa=\left.\frac{1}{2}\,\frac{\left|\partial_{\rho}g_{tt}\right|}{\sqrt{-g_{tt}\,g_{\rho\rho}}}\right|_
{\rho=\rho_{s}}=\frac{d-3}{2\,\rho_{s}}\(1-\gamma_{d}\,\lambda'\)~,
\ee where \be \left. \gamma_d \equiv
\rho_s^2\(\epsilon-\mu\)\right|_ {\rho=\rho_{s}}\label{def_gamma}~,\ee
and \be\lambda'\equiv \frac{\lambda}{\rho_{s}^{2}}.\ee $\lambda'$ is
the small dimensionless expansion parameter since the $\alpha'$
corrections can be trusted only when the string scale is small
compared to the black hole size. The values of $\gamma_d$ for
various dimensions are given in appendix A.
To leading order in $\lambda'$ we then have \be
\beta=\frac{2\,\pi}{\kappa}=\frac{4\,\pi\,\rho_{s}}{d-3}\(1+\gamma_{d}\,\lambda'\).
\label{periodS} \ee

The action is apparently ``scheme'' dependent.
Different actions can be obtained by redefinitions of the fields,
which are in this case the dilaton and the metric. Callan, Myers and
Perry~\cite{CMP}
introduced a different scheme by using field redefinitions of order $\lambda$
to eliminate higher derivative dilaton terms that appear after
a conformal transformation to the Einstein frame,
\be g_{\alpha\beta}\rightarrow e^{\frac{4}{d-2}\,\phi}g_{\alpha\beta}~.\ee
We will refer to the scheme adopted in the Einstein frame as ``the Einstein
scheme" and to the previous one as ``the String scheme.'' In the
Einstein scheme physical quantities are usually obtained and
interpreted in a simpler way. However, later in this paper, we would
like to perform a T-duality transformation on the
$\alpha'$-corrected metric. T-duality is expressed more naturally in
the string scheme and therefore we will adopt this scheme for most
of the paper.

In the Einstein scheme the action (\ref{Seff}) becomes~\footnote{There is also a
$(\nabla\phi)^4$ term in the Einstein frame which cannot be eliminated
without altering the string S-matrix~\cite{Tseytlin1};
we thank Arkady Tseytlin for pointing this out.
This term does not contribute to the solution in the linear approximation.}
\be
I_{eff}^E=
\frac{1}{16\,\pi G_d} \int d^{d}x
\,\sqrt{-g}\,\left(R-\frac{4}{d-2}\,(\nabla\phi)^{2}
+\frac{\lambda}{2}\,e^{-\frac{4}{d-2}\,\phi}\,R_{\mu\nu\rho\sigma}\,R^{\mu\nu\rho\sigma}
\right). \ee
The solution in this scheme is given by eq. (\ref{Scwa corrected}),
with $f$ given by \be
f(\rho)=1-\frac{\rho_{E}^{d-3}}{\rho^{d-3}}, \ee
where $\rho_E$ is the location of the
horizon in this scheme.
The corrections are
expressed by somewhat simpler expressions: \bea
\mu&=&-\epsilon, \\
\epsilon&=&\frac{(d-3)(d-4)}{4}\,\frac{\rho_{E}^{
d-5}}{\rho^{d-1}}\,\frac{\rho^{d-1}-\rho_{E}^{d-1}}{\rho^{d-3}-\rho_{E}^{d-3}},
\eea and the solution for $\varphi$ is the same expression as in the string
scheme $(\ref{varphisolution})$ with $\rho_E$ instead of $\rho_s$.

The periodicity of the Euclidean time is \be \label{periodE}
\beta=\frac{4\,\pi\rho_{E}}{d-3}\(1+\frac{(d-1)(d-4)}{2}\lambda'\).
\ee
Since the periodicity is invariant under field redefinitions we can
find from comparison of eqs. (\ref{periodS}) and (\ref{periodE}) a
relation between the location of the horizons in the two schemes:
\be
\rho_{E}=\rho_{s}\(1+\(\gamma_d-\frac{(d-1)(d-4)}{2}\)\lambda'\).
\label{translation} \ee The location of the horizon is the only free
dimensionfull parameter of the solution and as such it determines
all the physical quantities. Translation of this quantity between
the two schemes gives the translation of all the physical
quantities.

 As an example let us take the ADM mass in the Einstein
scheme~\cite{CMP},  \be
M=\frac{(d-2)\,\Omega_{d-2}\,\rho_{E}^{d-3}}{16\,\pi\,G_{d}}
\(1+\frac{(d-3)(d-4)}{2}\lambda'\).\ee The mass in the Einstein
scheme can be translated to the string scheme using
(\ref{translation}).

As a second example, we can apply this transformation to the free
energy that was calculated by Callan, Myers and Perry from the
Euclidean action in the Einstein scheme~\cite{CMP}, \be
\emph{F}_{E}=\frac{\Omega_{d-2}\,\rho_{E}^{d-3}}{16\,\pi\,G_{d}}\,\(1-\frac{d\,(d-3)}{2}\lambda'\),
\ee obtain it in the string scheme as \be
\emph{F}_{S}=\frac{\Omega_{d-2}\,\rho_{s}^{d-3}}{16\,\pi\,G_{d}}\,\(1+(d-3)\(\,\gamma_{d}-\frac{(d-2)^{2}}{2}\)\lambda'\),
\label{free energy}\ee and then derive the ADM mass in the string
scheme using the thermodynamic identity \be M=\frac{\partial \(\beta
\, F_{S}\)}{\partial
\beta}=\frac{(d-2)\,\Omega_{d-2}\,\rho_{s}^{d-3}}{16\,\pi\,G_{d}}
\(1+(d-3)\(\gamma_{d}-\frac{(d-2)(d-4)}{2}\)\lambda'\)~.\ee
The entropy in the string scheme is
 \be
 S=\beta\,\(M-F_{S}\)=
 \frac{\Omega_{d-2}\,\rho_{s}^{d-2}}{4\,G_{d}}\(1-\frac{(d-2)}{2}\(d^2-7d+10 - 2\gamma_d\)\lambda'\), \label{CMP entropy}
\ee and in the Einstein scheme \be
 S=\frac{\Omega_{d-2}\,\rho_{E}^{d-2}}{4\,G_{d}}\(1+(d-2)(d-3)\lambda'\).
\ee
The entropy can be written in terms of the ADM mass, which is a
scheme invariant quantity, and $\lambda'$ as
 \be S=2^{\frac{2d-3}{d-3}}\,\left[G_{d}\,\Gamma\(\frac{d-1}{2}\)\right]^{\frac{1}{d-3}}\,\sqrt{\pi}\(\frac{\,M}{d-2}\)^{\frac{d-2}{d-3}}\(1+\frac{(d-2)^{2}}{2}\lambda'\). \label{CMP
entropy inv}\ee

\section{The $\alpha'$ corrections to the $p_{L}=p_{R}$ solution}
\label{one correction section} In this section we calculate the leading
$\alpha'$ corrections to the case when $p \equiv p_{L}=p_{R}$ in
(\ref{hor-pol_metric}), namely, no winding charges:
$w=\alpha_{w}=0$. {}From now on we shall work in the string scheme
(unless otherwise specified). This solution is obtained from the
Schwarzschild-Tanghelini solution by adding a spectator
coordinate $x$ to the $d$-dimensional metric (to create a
$d+1$-dimensional black string), boosting along this direction and
then reducing to $d$ dimensions. Thus we lift the $d$-dimensional
solution (\ref{Scwa corrected}) to $d+1$ dimensions and write
$g_{xx}$ to order $\lambda'$ in the form
\be g_{xx}=1+\lambda'\,\xi(\rho)~.\ee
Then the equations of motion with the same appropriate boundary
conditions~\cite{CMP} as in the previous section give $\xi = 0$ as
the only solution. Thus the boosted solution is given by
(\ref{boosted})
 and after reduction to $d$ dimensions we get to first order in
 $\lambda'$:
\bea g_{tt}&=&\frac{g_{tt}}{\cosh^{2}(\alpha_{n})
+\sinh^{2}(\alpha_{n})\,g_{tt}}
=-\frac{f}{\Delta\left(\alpha_n\right)} \left(1+2\lambda'\mu\left(\rho\right)\frac{\rho_{s}^{2}\cosh^2(\alpha_n)}{\Delta\left(\alpha_n\right)}\right), \nonumber\\
\emph{A}_{t}^{n}&=&\frac{\sinh(\alpha_{n})\,
\cosh(\alpha_{n})\,(g_{tt}+1)}{\cosh^{2}(\alpha_{n})
+\sinh^{2}(\alpha_{n})\,g_{tt}}=\frac{1}{2}\left(\frac{\rho_{s}}{\rho}\right)^{d-3}\frac{\sinh2(\alpha_n)}
{\Delta\left(\alpha_n\right)}
\left(1-\frac{2\lambda'\mu\left(\rho\right)\,\rho^{d-3}\,f}{\rho_{s}^{d-5}\,\Delta\left(\alpha_n\right)}\right),
\nonumber\\
e^{-2\,\phi}&=&e^{-2\,\phi_0}\Big(1-2\,\lambda'
\varphi(\rho)\rho_s^2\Big)\(\cosh^{2}(\alpha_{n})
+\sinh^{2}(\alpha_{n})\,g_{tt}\)^{\frac{1}{2}}\label{gtt in string}\\
&&=e^{-2\,\phi_0}\,\sqrt{\Delta(\alpha_{n})}\,\(1-2\lambda'\varphi(\rho)\rho_s^2-\frac{\lambda'\,\rho_{s}^{2}\,\mu(\rho)\,f\,\sinh^{2}\alpha_{n}}{\Delta(\alpha_{n})}\)
\nonumber , \eea and the rest of the components of the metric remain
unchanged.

 The charge $p$ receives $\alpha'$ corrections:
 \be
 p=\frac{\left(d-3\right)\Omega_{d-2}\rho_s^{d-3}}{32\pi G_d}
\sinh(2\alpha_n)\(1+c_{d}\lambda'\)
 \label{one p in s}.\ee
 On the other hand, the chemical potentials (\ref{phi})
 are not affected by $\alpha'$
 corrections since the corrections to the vector potential
 vanish at the horizon. We will denote them by $\Phi \equiv
 \Phi_{L}=\Phi_{R}$. The chemical potential represents the response of the mass (energy) to a
change in the charge. When we introduce $\lambda'$-corrections we
change the ratio of mass to charge when the response is held fixed.

 The periodicity of the Euclidean time is
 \be
\beta = \frac{1}{T} = \frac{4\pi \rho_s\cosh(\alpha_n)}{d-3}
\left(1+\gamma_d\,\lambda'\right). \label{period boosted once}
 \ee
 This result can be interpreted as the boosted inverse temperature
 of the \Schw solution with the $\alpha'$ correction
 (\ref{periodS}). Note that the boost parameter does not receive
 $\alpha'$ corrections and hence it is kept in the same form in any scheme.

Since the free energy~(\ref{free energy}) does not change under
boost, the ADM mass can be derived from the thermodynamic identity
\be M=\frac{\partial\( \beta\,F_S\)}{\partial \, \beta}+2\,p\,\Phi.
\label{follow}\ee $\beta\,F$ is the value of the Euclidean action.
The ADM mass is
\bea M=\frac{(d-2)\,\Omega_{d-2}\,\rho_{s}^{d-3}}{16\,\pi\,G_{d}}
\Big[1&+&(d-3)\(\gamma_{d}-\frac{(d-2)(d-4)}{2}\)\lambda'\nonumber\\
&+&\(\frac{d-3}{d-2}\)\,
\sinh^{2}(\alpha_n)\(1+c_{d}\lambda'\)\Big]. \label{Mass in s}\eea
The charge $p$ and the mass $M$ are changed as a result of the boost
when $\rho_{s}$ and $\lambda'$ are held fixed. This way the
uncharged \Schw solution is transformed to a charged solution. Note
that the ratio
\be \frac{\Delta \, M}{\Delta \, p}=\tanh(\alpha_{n}),\ee
where $\Delta \,M$ and $\Delta \, p$ are the change in the mass and the charge,
is free of $\alpha'$ corrections.

One can obtain the corrected entropy from the thermodynamics as
well: \be
S=\beta\(M-F_S-2\,p\,\Phi\)=\frac{\Omega_{d-2}\,\rho_{s}^{d-2}}{4\,G_{d}}\,\cosh(\alpha_{n})\(1-\frac{(d-2)}{2}\(d^2-7d+10
- 2\gamma_d\)\lambda'\). \label{boosted entropy} \ee This result is
the Callan-Myers-Perry entropy (\ref{CMP entropy}) boosted with the
parameter $\alpha_{n}$. Before the reduction the entropy is the
product of the area of the black string times a function of
$\lambda'$, and when we boost along the compact dimension we expect
to get a boost factor of $\cosh(\alpha_{n})$ which reflects the
change in its size.

It is useful to write the parameters $\rho_s$ and $\alpha_n$ in terms
of $M$ and $p$, i.e. to invert the relations (\ref{one p in s}) and
(\ref{Mass in s}):
\bea
\sinh^2{\alpha_n}&=&\sinh^2{\alpha_n^0}\(1-\frac{2\,(d-2)^2}{d-2+\tanh^2\alpha_n^0}\lambda'\)~,
\label{conv1}\\
\rho_s&=&\rho_0\(1+\left[\frac{(d-2)^2\(1+\tanh^2(\alpha_n^0)\)}
{(d-3)\(d-2+\tanh^2(\alpha_n^0)\)}-\frac{c_d}{d-3}\right]\lambda'\)~,
\label{conv2}\eea
where $\alpha_{n}^0$ and $\rho_0$ are the values of $\alpha_n$ and
$\rho_s$ when $\lambda'=0$ and are given in terms of the mass and
the charge in eqs. (\ref{seder01},\ref{seder02}).

Using (\ref{conv1},\ref{conv2}) we can express the entropy
(\ref{boosted entropy}) in terms of scheme invariant quantities --
the mass $M$ and the charge $p$: \be
S=S_{\lambda'=0}\(M,p\)\(1+\frac{(d-2)^2}{2}\,\lambda'\),\label{boosted
entropy inv}\ee where $S_{\lambda'=0}$ is given in eq.
(\ref{entroppy_base}) where it was explicitly expressed in terms of
$p$ and $M$.

We see that the form of the perturbative $\lambda'$ correction to
the entropy is a multiplicative factor which we encountered in the
Callan-Myers-Perry solution (\ref{CMP entropy inv}).~\footnote{This
result is general, as we shall see later.} Note that the higher
derivative correction increases the value of the entropy and does not
depend on the charge.
Moreover, in the extremal limit, $S\rightarrow 0$
even with the perturbative $\lambda'$ correction.

In a similar manner we can express the temperature as a function of
$M$ and $p$: \be
T=T_{\lambda'=0}(M,p)\left[1+\frac{d-2}{2}\(1-\frac{(d-3)^{2}}{\delta}\)\lambda'\right],
\ee where $T_{\lambda'=0}(M,p)$ and $\delta$ are given in
(\ref{temperature_base}) and (\ref{sdelta}), respectively.
Note that in $d=4$ the correction increases the temperature and for
$d>4$ it is decreased.

Finally, let us inspect the \emph{mass to charge ratio}.
We can compute the higher derivative
corrections to the mass to charge ratio using the expressions that
we already obtained for the corrected mass and charge -- (\ref{one p
in s}) and (\ref{Mass in s}):
\bea
\frac{M}{p}=\tanh{(\alpha_n)}&+&\frac{d-2}{(d-3)\sinh(\alpha_n)\cosh(\alpha_n)}
\nonumber\\
&+&\frac{\lambda'(d-2)\big[2(d-3)\gamma_d-2\,c_d-(d-2)(d-3)(d-4)\big]}{2(d-3)\sinh(\alpha_n)\cosh(\alpha_n)}~.
\eea
Using the relation between the constants $\gamma_d$ and $c_d$
(see (\ref{constants}) in appendix A),
one can write the following expression for the
ratio: \be
\frac{M}{p}=\tanh{(\alpha_n)}+\frac{d-2}{(d-3)\sinh(\alpha_n)\cosh(\alpha_n)}\Big[1-\lambda'(\,
d-2)\Big]~. \ee
We see that the corrections decrease the mass to charge ratio for
any dimension. The effect is stronger as we go to a higher dimension
and is minimal for $d=4$.

\section{The $\alpha'$ corrections for general $(p_{L},p_{R})$}

In this section we shall add winding charge by first T-dualizing the
solution of the previous section, in order to change the fundamental
string momentum into winding, and then turning on another boost to
add momentum charge again, as was explained in section~\ref{first
section}. The difference here is that we use the CMP solution
described in section~\ref{CMP} instead of the \Schw solution, and
$\alpha'$-corrected T-duality rules.

So first, we should discuss the $\alpha'$-corrected T-duality
in the $x$ direction of the solution in the previous section. When
one includes the $\alpha'$ corrections in the action the T-duality
rules~(\ref{Trules}) are modified \cite{Tseytlin2,Meissner,Kaloper}.  In
\cite{Meissner} a different scheme is introduced in which the
$R_{\mu\nu\rho\sigma}\,R^{\mu\nu\rho\sigma}$ term in (\ref{Seff}) is
transformed to a Gauss-Bonnet term
\be I_{GB} \equiv R_{\mu\nu\rho\sigma}\,
R^{\mu\nu\rho\sigma}-4\,R_{\mu\nu}\,R^{\mu\nu}+R^{2}~,\label{gaussbonnet}\ee
as well as
terms that involve derivatives of the scalar field. The full action
in this scheme is given in  eq. (4.5) of~\cite{Meissner}. This
``Gauss-Bonnet scheme" is obtained from the string scheme
(\ref{Seff}) using the following field redefinitions,~\footnote{The
field redefinitions are written for the case when the antisymmetric
tensor vanishes. Otherwise there are some additional terms.} \bea
g_{\mu\nu} \rightarrow g_{\mu\nu}+2\,\lambda\,R_{\mu\nu}, \quad
\quad \phi \rightarrow \phi +
\frac{\lambda}{4}\,R-\lambda\,\(\partial\,\phi\)^{2}.
 \eea
The Callan-Myers-Perry solution does not change under these field
redefinitions since all the terms of order $\alpha'$ are
identically zero.

In this scheme, Kaloper and Meissner \cite{Kaloper} obtained the
$\alpha'$ corrected rules, written here for our particular case of a
diagonal metric which depends only on one coordinate $\rho$ and
boosted along an additional direction $x$:
 \bea
g^{T}_{tt}&=&g_{tt}-\frac{g_{xt}^{2}}{g_{xx}},
\label{gggtt}\\
g^{T}_{xx}&=&\frac{1}{g_{xx}}\(1+\frac{\lambda\,\(g_{xx,\rho}\)^{2}}{g_{xx}^{2}\,g_{\rho\rho}}+\frac{\lambda\,g^{tt}\,g_{xx}^{2}\,\(\partial_{\rho}\,V\)^{2}}{g_{\rho\rho}}\),\\
B^{T}_{xt}&=&\frac{g_{xt}}{g_{xx}}-\frac{\lambda\,\partial_{\rho}\,V\,g_{xx,\rho}}{g_{\rho\rho}\,g_{xx}},\\
\phi^T&=&\phi+\frac{1}{4}\,\ln\(\frac{g^T_{xx}}{g_{xx}}\),
\label{ppp} \eea where
\be V \equiv \frac{g_{xt}}{g_{xx}}~.\ee

We are now ready to apply the solution generating procedure.
We first perform a boost (\ref{boosted}) on the CMP solution
(\ref{Scwa corrected}) with $\alpha_{w}$ as a boost parameter which
can be interpreted as related to the winding modes after a
subsequent T-duality (\ref{gggtt}) -- (\ref{ppp}): \bea
g^{T,\alpha_w}_{tt}&=&-\frac{f}{\Delta(\alpha_w)}\Big[1+\frac{2\lambda\,\mu(\rho)\,\cosh^{2}\alpha_{w}}{\Delta(\alpha_w)}\Big],\\
g^{T,\alpha_w}_{xx}&=&\frac{1}{\Delta(\alpha_w)}\Big[1+\frac{2\,\lambda\,\mu(\rho)\,f\,\sinh^2\alpha_w}{\Delta(\alpha_w)}-\frac{\lambda\,(d-3)^2\rho_s^{2\,(d-3)}\sinh^2\alpha_{w}^2}{\rho^{2\,(d-2)}\Delta(\alpha_w)}\Big],\\
B^{T,\alpha_w}_{xt}&=&\frac{1}{2}\left(\frac{\rho_s}{\rho}\right)^{d-3}\frac{\sinh2\alpha_w}
{\Delta\left(\alpha_w\right)}\Big[1-\frac{2\lambda\,\mu(\rho)\rho_{s}^{d-3}f}{\rho^{d-3}\Delta(\alpha_w)}-\frac{\lambda\,(d-3)^2\,f\,\rho^{d-3}_s}{\rho^{d-1}\,\Delta(\alpha_w)^2}\Big],
\label{last term}\\
\phi^{T,\alpha_w}&=&\phi_0-\frac{1}{2}\,\ln(\Delta(\alpha_w))
\nonumber
\\
&+&\lambda\Big[1+\varphi(\rho)+\frac{\mu(\rho)\,f\,\sinh^2\alpha_w}{\Delta(\alpha_w)}-\frac{(d-3)^2\rho_{s}^{2(d-3)}\sinh^2\alpha_w}{4\,\rho^{2\,(d-2)}\Delta(\alpha_w)}\Big],
\eea where $\Delta(x),\mu(\rho),\varphi(\rho)$ are given in
(\ref{delta}), (\ref{Scwa corrected})--(\ref{endmu}).
The rest of the components remain unchanged.

 Now we perform the
second boost in the $x$ direction with a boost parameter $\alpha_n$, and after
reduction to $d$ dimensions we find:
 \bea
g_{tt}&=&\frac{g^{T,\alpha_w}_{xx}\,g^{T,\alpha_w}_{tt}}
{\cosh^{2}(\alpha_{n})\,g^{T,\alpha_w}_{xx}+\sinh^{2}(\alpha_{n})\,g^{T,\alpha_w}_{tt}}
\label{gtthat}\\
&=&-\frac{f}{\Delta(\alpha_n)\,\Delta(\alpha_w)}\left[1+\frac{2\,\lambda'\,\rho^2_s\,\mu(\rho)}{\Delta(\alpha_n)\,\Delta(\alpha_w)}\,
-\frac{2\,\lambda'\,\mu(\rho)\,\rho^{2(d-3)}\,
\sinh^{2}(\alpha_{n})\,\sinh^{2}(\alpha_{w})}{\rho_s^{2(d-4)}\Delta(\alpha_n)\,\Delta(\alpha_w)}\right.
\nonumber\\
&+&2\lambda'\rho_s^2\mu(\rho)\(\frac{\sinh^{2}\alpha_n}{\Delta(\alpha_n)}+\frac{\sinh^{2}\alpha_w}{\Delta(\alpha_w)}\)+\left.
\frac{(d-3)^2\,\lambda'\,\rho_s^{2(d-2)}f\sinh^2(\alpha_n)\sinh^2(\alpha_w)}{\rho^{2(d-2)}\Delta(\alpha_n)\,\Delta(\alpha_w)}\right],
\nonumber \\
g_{\rho\rho}&=&f^{-1}\,\left( 1+2\,\lambda'\rho_s^2\,\epsilon(\rho)\right),\\
\emph{A}_{t}^{n}&=&\frac{\sinh(\alpha_{n})\,\cosh(\alpha_{n})
\(g^{T,\alpha_w}_{xx}+g^{T,\alpha_w}_{tt}\)}{\cosh^{2}(\alpha_{n})\,
g^{T,\alpha_w}_{xx}+\sinh^{2}(\alpha_{n})\,g^{T,\alpha_w}_{tt}}\nonumber\\
&=&\frac{1}{2}\left(\frac{\rho_s}{\rho}\right)^{d-3}\frac{\sinh2\alpha_n}
{\Delta\left(\alpha_n\right)}\Big[1-\frac{2\,\lambda'\,\mu(\rho)\,\rho^{d-3}\,f}
{\rho_s^{d-5}\Delta(\alpha_n)}-\frac{\lambda'\,
(d-3)^2\,\rho_s^{d-1}\,f\,\sinh^2\alpha_w}{\rho^{d-1}\Delta(\alpha_n)\,\Delta(\alpha_w)}\Big],\\
\emph{A}^{w}_{t}&=&B^{T,\alpha_w}_{xt}, \label{awthat}\\
e^{-2\,\phi}&=&e^{-2\,\phi_0}\,\sqrt{\Delta(\alpha_n)\,\Delta(\alpha_w)}\,
\left[1-2\,\lambda'\,\rho_s^2\,\varphi(\rho)-\lambda'\rho_s^2\mu(\rho)f\(\frac{\sinh^2\alpha_n}{\Delta(\alpha_n)}
+\frac{\sinh^2\alpha_w}{\Delta(\alpha_w)}\)\right.\nonumber\\
&-&\left.\lambda'\frac{(d-3)^2\,\rho_s^{2(d-2)}\,f\sinh^2\alpha_n\sinh^2\alpha_w}{2\rho^{2\,(d-2)}\Delta(\alpha_w)\Delta(\alpha_n)}\right],
\label{ephihat} \eea
where $f$ is given in (\ref{fffff}), and we used the dimensionless expansion
parameter $\lambda'=\frac{\lambda}{\rho_s^2}$ instead of $\lambda$.

Note that the metric components and the dilaton with the $\alpha'$
corrections are invariant under the exchange of momentum and winding,
$\alpha_n \leftrightarrow \alpha_w$.
This is a manifestation of the T-duality
being a symmetry even at the level of first order corrections to the
low energy effective action.

The horizon is located at $\rho=\rho_s$, and the Euclidean time has
a period \be \beta = \frac{1}{T} = \frac{4\pi
\rho_s\cosh(\alpha_n)\cosh(\alpha_w)}{d-3}
\left(1+\gamma_d\,\lambda'\right). \label{period boosted twice}
 \ee The charges can be read from the asymptotic behavior of the
vector potentials: \be \label{general p}
 p_{R/L}=\frac{\left(d-3\right)\Omega_{d-2}\rho_s^{d-3}}{32\pi G_d}
\(\sinh2\alpha_n \pm \sinh2\alpha_w\)\(1+c_{d}\lambda'\),\ee
and the chemical potentials~(\ref{phi}) again do not receive $\alpha'$ corrections.

Since the integrand of the Euclidean action is invariant under
T-duality, and following the discussion near eq. (\ref{follow}),
the free energy (\ref{free energy}) is valid in the present case as
well. The mass is thus \bea M=
\frac{(d-2)\,\Omega_{d-2}\,\rho_{s}^{d-3}}{16\,\pi\,G_{d}}\Big[1+(d-3)\(\gamma_{d}-\frac{(d-2)(d-4)}{2}\)\lambda'
\nonumber
\\+\(\frac{d-3}{d-2}\)\,
\(\sinh^{2}\alpha_n+\sinh^{2}\alpha_w\)\(1+c_{d}\lambda'\)\Big],
\label{general m}\eea and the entropy is
 \bea
S&=&\beta\(M-F_S-p_L\,\Phi_L-p_R\,\Phi_R\) \nonumber \\
&=&\frac{\Omega_{d-2}\,\rho_{s}^{d-2}}{4\,G_{d}}\,\cosh\alpha_{n}\,\cosh\alpha_{w}\(1-\frac{(d-2)}{2}\(d^2-7d+10
- 2\gamma_d\)\lambda'\). \label{pfe} \eea One can show that \be
S=S_{\lambda'=0}(M;p_L,p_R)\left(1+\frac{(d-2)^2}{2}\lambda'\right)~,
\label{full entropy} \ee where $S_{\lambda'=0}(M;p_L,p_R)$ is the
value of the leading order entropy -- the Bekenstein-Hawking one --
of a black fundamental string with mass $M$ and charges $(p_L,p_R)$.
The derivation of (\ref{full entropy}) for the general case is
sketched in appendix B.

Another useful form to present the entropy is to write it with
the horizon radius in the Einstein scheme, given in eq.
(\ref{translation}): \be
S=\frac{\Omega_{d-2}\,\rho_E^{(d-2)}}{4\,G_{d}}\,\cosh\alpha_{n}\,\cosh\alpha_{w}\(1+(d-2)(d-3)\lambda'\)~.
\label{einstein entropy}\ee
The analysis of this section reveals, in particular,
that the numerical factor in front of the $\alpha'$ correction term
to the entropy is independent of the value of the charges.

A special case is when, say, $p_R\rightarrow M$ for any $p_L\leq M$.
In the heterotic string this corresponds to BPS fundamental strings.
However, as discussed below eq. (\ref{invttt}), in this limit, in particular
$\rho_s\rightarrow 0$ and, correspondingly, the $\lambda'$ expansion is not valid.
Nevertheless, near extremal solutions are valid as long as the black hole
size is much bigger than the string length scale, $\rho_s^2\gg\alpha'$.

Finally, let us inspect the \emph{mass to charge ratio}.
Using eqs. (\ref{general p}), (\ref{general m}) and the relation of the
constants (\ref{constants}) in appendix A, we find that
\be
\frac{M}{p_{R/L}}=\frac{d-2+(d-3)\(\sinh^2\alpha_n+\sinh^2\alpha_w\)-\lambda'(d-2)^2}{\(d-3\)\left[\sinh\alpha_n\cosh\alpha_n\pm\sinh\alpha_w\cosh\alpha_w\right]}.
\ee
This shows the same behavior as in the case of a single
charge. The $\alpha'$ correction tends to decrease the mass to
charge ratio for any dimension and any value of the charges.

\section{Calculation of the Noether charge entropy}

The entropy was derived in the previous sections in the Euclidean
approach. Wald derived a formula realizing that the entropy is the
Noether charge for diffeomorphism invariant
lagrangians~\cite{wald1}. Wald's formula is consistent with the
first law of thermodynamics and therefore also with the Euclidean
approach. It provides an alternative way of computing the
entropy directly from the action, and will allow us to express
the entropy for the {\it exact} solution to the four derivative effective action
in terms of the black hole area.

The Noether charge entropy is given in a form of an integral over
fields on a spatial section of the horizon $\Sigma$. For theories
without derivatives of the Riemann tensor Wald's formula takes the
following form~\cite{myers}: \be
S_{BH}=-2\,\pi\,\int_{\Sigma}\frac{\partial\,L}{\partial R_{\mu \nu
\rho \sigma}}\,\epsilon_{\mu\nu}\,\epsilon_{\rho\sigma}\,
\sqrt{h}\,d\Omega_{d-2}~,\ee where the action of the $d$-dimensional
theory is
\be I=\int\!d^{d}x\sqrt{-g}\,L~,\ee
and $\epsilon_{\mu\nu}$ is the binormal to the spatial section of the
horizon $\Sigma$ -- the volume element orthogonal to it. The binormal
is given by $\epsilon_{\mu\nu}=\nabla_{\mu}\chi_{\nu}$,
where $\chi_{\nu}$ is a Killing field normalized so that
$\epsilon_{\mu\nu}\epsilon^{\mu\nu}=-2$. $\sqrt{h}\,d\Omega_{d-2}$
is the volume element induced on $\Sigma$.

The low energy
effective action of string theory with leading order $\alpha'$
corrections is given by~\cite{Tseytlin1}
\bea &&I_{eff}^{\lambda}=
\frac{1}{16\,\pi\,G_{d}} \int d^{d}x
\,\sqrt{-g}\,e^{-2\,\phi}\,\left(R+4\,(\nabla\phi)^{2}-\frac{1}{12}\,H^{2}
\right. \nonumber \\
&&+\frac{\lambda}{2}\,\left[I_{GB}-I_{R\phi}-\frac{1}{2}\,I_{RH}+
16\,\nabla^{2}\phi\,\(\nabla\phi\)^{2}-16\,\(\nabla\phi\)^{4}+\frac{2}{3}\,H^{2}\,\(\nabla
\phi\)^{2}-\frac{1}{8}\,H_{\mu\nu}^{2}\,H^{2\mu\nu}
\right.\nonumber \\
&&
\left.\left.+2\,\(\nabla^{\mu}\nabla^{\nu}\phi\,H_{\mu\nu}^{2}-\frac{1}{3}\,\nabla^{2}\phi\,H^{2}\)
-\frac{1}{24}\,H_{\mu\nu\lambda}\,H^{\nu}_{\rho\alpha}\,H^{\rho\,\sigma\,\lambda}\,H_{\sigma}^{\mu\alpha}
-\frac{1}{144}\(H^{2}\)^{2} \right]\right)~, \label{iefflambda}\eea
where $I_{GB}$ is given in (\ref{gaussbonnet}),
\bea
I_{R\phi}&=&16\,\(R^{\mu\nu}-\frac{1}{2}\,g^{\mu\nu}\,R\)\nabla_{\mu}\phi\nabla_{\nu}\phi~, \\
I_{RH}&=&\(R_{\mu\nu\lambda\rho}\,H^{\mu\nu\alpha}\,H^{\lambda\rho}_{\alpha}-2\,R^{\mu\nu}\,H_{\mu\nu}^{2}
+\frac{1}{3}\,R\,H^{2}\)~,
\eea and \bea
H_{\mu\nu}^{2}&=&H_{\mu\alpha\beta}H_{\nu}^{\alpha\beta}~,\\
H^{2}&=&H_{\mu\alpha\beta}\,H^{\mu\alpha\beta}~.
 \eea
We use here the effective action
obtained by Jack and Jones~\cite{Jack} when we ignore the boundary
terms that do not contribute to the Noether charge entropy.

In order to apply Wald's formula to the action above we have to
carry out the variation of the action with respect to
$R_{\mu\nu\rho\sigma}$ regarding the Riemann tensor as formally
independent of the metric $g_{\mu\nu}$. For example, in the case of
Einstein-Hilbert lagrangian, \be L_{EH}=\frac{1}{16\,\pi\,G_{d}}R~, \ee
we obtain \be \frac{\delta L_{EH}}{\delta R_{\alpha \beta \gamma
\delta}}=\frac{1}{32\,\pi\,G_{d}}\(g^{\alpha\gamma}\,g^{\beta\delta}-g^{\beta\gamma}\,g^{\alpha\delta}\).
\ee Therefore, the terms in the $\alpha'$ corrections to the action
that may contribute to the entropy are the ones which involve the
Riemann tensor explicitly, namely, $I_{GB}$, $I_{R\phi}$ and
$I_{RH}$: \bea \frac{\delta I_{GB}}{\delta R_{\alpha \beta \gamma
\delta}}&=&2\,R^{\alpha\beta\gamma\delta}-4\,\(g^{\alpha\gamma}\,R^{\beta\delta}-g^{\beta\gamma}\,R^{\alpha\delta}\)
+R\,\(g^{\alpha\gamma}\,g^{\beta \delta}-g^{\beta \gamma}\,g^{\alpha\delta}\),\\
\frac{\delta I_{R\phi}}{\delta R_{\alpha \beta
\gamma\delta}}&=&8\,\(\nabla^{\alpha}\phi\,\nabla^{\gamma}\phi\,g^{\beta\delta}-g^{\beta\gamma}\,\nabla^{\alpha}\phi\,\nabla^{\delta}\phi\)-
4\,\nabla_{\mu}\phi\,\nabla^{\mu}\phi\,\(g^{\alpha\gamma}\,g^{\beta\delta}-g^{\alpha\delta}\,g^{\beta\gamma}\), \\
\frac{\delta I_{RH}}{\delta R_{\alpha \beta \gamma\delta}}&=&
H^{\alpha\beta\mu}H^{\gamma\delta}_{\mu}-\left[(H^{\alpha\gamma})^{2}\,
g^{\beta\delta}-g^{\beta\gamma}\,(H^{\alpha\delta})^{2}\right]
+\frac{1}{6}H^{2}\(g^{\alpha\gamma}\,g^{\beta\delta}-g^{\beta\gamma}\,g^{\alpha\delta}\), \eea
where a useful formula for the computation of the above results is that for
 $I=R^{\alpha\beta}\,K_{\alpha\beta}$
when $K_{\mu\nu}$ is a symmetric tensor,
\be
\frac{\delta\,I}{\delta R_{\alpha\beta\gamma\delta}}=\frac{1}{2}\,
\(g^{\alpha\gamma}\,K^{\beta\delta}-g^{\alpha\beta}\,K^{\gamma\delta}\).
\ee
Let us consider this action as the action before KK reduction,
namely in $d+1$ dimensions. Since the value of the entropy does not
change by a KK reduction we can inspect it with an additional
dimension and then write it in terms of $d$-dimensional quantities.

{}For a static metric a Killing field is $\frac{\partial}{\partial t}$.
{}For a static metric of the form \be
ds^{2}=-g_{tt}\,dt^{2}+g_{\rho\rho}\,d\rho^{2}+g_{ab}\,dx^{a}\,dx^{b}~,
\label{metricform}\ee the binormal to the horizon is given by
$\epsilon_{\rho t}=\sqrt{-g_{tt}\,g_{\rho\rho}}.$
Then we are left with variations only with respect to $R_{\rho t \rho t}$. We
show now that for
\begin{itemize}
\item A metric of the form (\ref{metricform})
\item A scalar field $\phi=\phi(\rho)$ (depends only on the radial
coordinate) and it is regular at the horizon $\rho=\rho_{h}$
\item The only component of $H_{\alpha\beta\gamma}$ which does not
vanish is $H_{\rho tx}$ (and permutation of the indices)
\end{itemize}
the variations of $I_{R\phi}$ and $I_{RH}$ do not contribute to the
entropy.

With the above assumptions we get that \be
\epsilon_{\alpha\beta}\epsilon_{\gamma\delta}\,\frac{\delta
I_{R\phi}}{\delta
R_{\alpha\beta\gamma\delta}}=4\,\epsilon_{\rho t}^{2}\,\frac{\delta
I_{R\phi}}{\delta R_{\rho t \rho t}}\propto g^{\rho\rho}\(\partial_{\rho}\phi\)^{2}.
\ee Since the horizon for this type of metric is defined as the
surface where $g^{\rho\rho}=0$, the term $I_{R\phi}$ does not contribute
to the entropy. Substitution of a metric of the form
(\ref{metricform}) in the expression for $\frac{\delta
I_{RH}}{\delta R_{\rho t\rho t}}$ gives zero identically.

The terms that contain the additional fields to the metric do not
contribute to the entropy in $d+1$ dimensions and hence do not
contribute to the entropy in $d$ dimensions as well. The only
contribution to the correction comes from the Gauss-Bonnet term
$I_{GB}$ that gives after implementation of Gauss-Codazzi equations
a term which is proportional to the scalar curvature of the sphere
$S^{d-2}$ (see for example~\cite{wald2} and a calculation using
differential forms in~\cite{Ross}).

The final expression for the entropy in $d$ dimensions in the string
scheme is thus \be
S=\frac{A_H}{4\,G_{d}}\,e^{-2\,\phi_{h}}\(1+\lambda'\,R_{S^{d-2}}\)=\\
\frac{A_H}{4\,G_{d}}\,e^{-2\,\phi_{h}}\(1+\lambda'\,(d-2)\,(d-3)\)~,
\ee where $\lambda'=\frac{\lambda}{\rho_{h}^{2}}$, $\rho_{h}$ is the
radius of the horizon and $A_H$ is its area. In the Einstein frame
we obtain
\be S=\frac{A_H}{4\,G_{d}}\,\(1+\lambda'\,(d-2)\,(d-3)\)~.\ee

Two comments are in order:
\begin{itemize}

\item
The formula for the entropy derived from the action (\ref{iefflambda}) is valid not only
for a preturbative solution in $\alpha'$  but also for the exact solution for
this action. The only difference will appear in a different expression for
the area of the black hole. In this sense this result is more general than the
perturbative result in $\alpha'$ (\ref{einstein entropy}).

\item
The entropy depends only on the area (and the value of the scalar field at the
horizon in the string scheme).
The correction to the area is a charge independent multiplicative factor
and is in agreement with the calculation in the Euclidean approach (\ref{einstein entropy}).
The charges thus enter the expression for the entropy only
through their effect on the area.

\end{itemize}

To recapitulate,
we found that the correction to the entropy depends on the charges
only through the horizon area. Previously, it was observed
in~\cite{GKRS} that the entropy of two dimensional charged black
holes is proportional to the area of the horizon for any value of
the charges and the mass. The entropy rather than being a more
general function of the charges keeps the dependence on the charges
only through the horizon area. An explanation to this phenomenon
appeared in~\cite{Brustein}. The entropy of higher derivative
theories can be interpreted as the area of the horizon in units of
the effective gravitational coupling. The effective gravitational
coupling is computed from the effective kinetic term for metric
perturbations on the horizon. This can be further interpreted as the
effective gravitational coupling for graviton exchange. Hence,
the electric charges do not affect such interactions.

\bigskip
\noindent{\bf Note Added:}
Exact numerical analysis in various types of four derivative actions
was done e.g. in~\cite{ohta} and references therein;
we thank Nobuyoshi Ohta for pointing this out to us.
It will be interesting to extend our study to these solutions.

\bigskip
\noindent{\bf Acknowledgements:}
We thank A.~Tseytlin for his comments.
This work was supported in part
by the BSF -- American-Israel Bi-National Science Foundation,
by a center of excellence supported by the Israel Science Foundation
(grant number 1468/06), DIP grant H.52, and the Einstein Center at the Hebrew University.
AG thanks the EFI at the University of Chicago for
its warm hospitality during the final stages of writing this work.
DG thanks the National Sciences and Engineering
Research Council of Canada for the financial support.

\appendix

\section{Expressions}

In this appendix we present some expressions used in the text;
they are taken from \cite{CMP}.
The function $K_{d}(x)$ for even $d$ is given by \bea
K_{d}(x)&=&-\frac{\pi}{2\,(d-3)}\,\tan\(\frac{\pi}{d-3}\)-\ln(x)\nonumber\\
&+&\sum_{l=1}^{\frac{d-4}{2}}\cos\(\frac{d-5}{d-3}\,(2\,l-1)\pi\)\ln\(1+2\,x\,\cos\(\frac{2\,l-1}{d-3}\,\pi\)+x^2\) \nonumber\\
&+&2\,\sum_{l=1}^{\frac{d-4}{2}}\sin\(\frac{d-5}{d-3}\,(2\,l-1)\pi\)\,
\arctan\(\frac{x+\cos\(\frac{2\,l-1}{d-3}\,\pi\)}{\sin\(\frac{2\,l-1}{d-3}\,\pi\)}\)~,
\eea
while for odd $d$
\bea K_{d}(x)&=&-\ln(x^2+x)-\sum_{l=1}^{\frac{d-5}{2}}\cos\(\frac{d-5}{d-3}\,2\,l\,\pi\)\ln\(1-2\,x\,\cos\(\frac{2\,l\,\pi}{d-3}\)+x^2\) \nonumber\\
&+&2\,\sum_{l=1}^{\frac{d-5}{2}}\sin\(\frac{d-5}{d-3}\,2\,l\,\pi\)\,
\arctan\(\frac{x+\cos\(\frac{2\,l}{d-3}\,\pi\)}{\sin\(\frac{2\,l}{d-3}\,\pi\)}\)~.
\eea
The constants $c_d$ and $\gamma_d$ (defined in eq. (\ref{def_gamma}))
for various dimensions are given by
\be
c_{d}=d-2-\frac{4\,(2d-3)}{(d-3)(d-1)}-\frac{2\,(d-2)}{d-3}K_{d}(1)-\frac{2\,(d-2)}{d-3}\ln(d-3)~,
\ee
\be
\gamma_d={c_d\over d-3}+\frac{(d-2)^2(d-5)}{2(d-3)}~. \label{constants}\ee
The numerical values of $\gamma_d$ and $c_d$ are given (up to three digits after the dot)
in table 1.
\begin{table}[htb]

\label{table:c_d and gamma_d} 
\setlength{\tabcolsep}{10pt}
\begin{tabular}{ l l l }
 \hline \hline
 $d$ & $\gamma_d$ & $c_d$
 \\
 \hline

 $4$  & $11/6$   & $23/6$   \\
 $5$  & $17/4$    & $17/2$      \\
 $6$  & 7.718   & 15.154   \\
 $7$  & 12.201  & 23.804   \\
 $8$  & 17.690  & 34.452   \\
 $9$  & 24.183  & 47.099   \\
 $10$ & 31.678  & 61.745   \\
 $11$ & 40.174  & 78.391   \\
 $12$ & 49.671  & 97.037   \\
 $13$ & 60.168  & 117.683  \\
 $14$ & 71.666  & 140.328  \\
 $15$ & 84.164  & 164.974  \\
 $16$ & 97.663  & 191.619  \\
 $17$ & 112.162 & 220.265  \\
 $18$ & 127.661 & 250.910  \\
 $19$ & 144.160 & 283.554  \\
 $20$ & 161.659 & 318.201  \\
 $21$ & 180.158 & 354.844  \\
 $22$ & 199.657 & 393.490  \\
 $23$ & 220.157 & 434.137  \\
 $24$ & 241.656 & 476.781  \\
 $25$ & 264.156 & 521.423  \\

 \hline\hline
    \end{tabular}
\caption{The numerical values of $\gamma_d$ and $c_d$ in various
dimensions.}

\end{table}

\section{The entropy for general $(p_{L},p_{R})$}

Here we show how to obtain eq. (\ref{full entropy}). We would like to write
the parameters $\rho_s$, $\alpha_{n}$ and $\alpha_{w}$ in terms of
$M,$ $p_{L}$ and $p_{R}$. Actually, it is not necessary to obtain
the inversion of (\ref{general p}) and (\ref{general m}) explicitly.
Instead, let us write: \bea
\sinh\alpha_{n}&=&\sinh\alpha_{n}^{0}\(1+a_{n}\,\lambda'\),\label{conversed1}\\
\sinh\alpha_{w}&=&\sinh\alpha_{w}^{0}\(1+a_{w}\,\lambda'\),\label{conversed2}\\
\rho_s&=&\rho_{0}\(1+b\,\lambda'\), \label{conversed3}\eea where we
denote by $\sinh\alpha_{n}^{0}$, $\sinh\alpha_{w}^{0}$ and
$\rho_{0}$ the inversion of (\ref{general p}) and (\ref{general m})
in the zeroth order (when $\lambda'=0$). Substitution into
(\ref{general p}) and (\ref{general m}) gives  \bea
   && a_n =
       - \frac{\(d-2\)^2 \(1+\tanh^2\alpha_{w}^{0}\)}
            { \left[d - 2 +\tanh^2\alpha_{w}^{0}+\tanh^2\alpha_{n}^{0}-(d-4)\tanh^2\alpha_{w}^{0}\tanh^2\alpha_{n}^{0}\right]}~,
  \nonumber  \\
  &&  a_w  =
        - \frac{\(d-2\)^2\(1+\tanh^2\alpha_{n}^{0}\)}
            { \left[d - 2 +\tanh^2\alpha_{w}^{0}+\tanh^2\alpha_{n}^{0}-(d-4)\tanh^2\alpha_{w}^{0}\tanh^2\alpha_{n}^{0}\right]}~,
   \\ \nonumber
  &&  b  =
       \frac{(d-2)^2\(1+\tanh^2{\alpha_{n}^{0}}+\tanh^2{\alpha_{w}^{0}}
       +\tanh^2{\alpha_{n}^{0}}\tanh^2{\alpha_{w}^{0}}\)}{(d-3)\left[d-2+\tanh^2{\alpha_{n}^{0}}
       +\tanh^2\alpha_{w}^{0}-(d-4)\tanh^2\alpha_{n}^{0}\tanh^2\alpha_{w}^{0}\right]}-\frac{c_d}{d-3}~.
\eea Thus, substituting eqs.
(\ref{conversed1},\ref{conversed2},\ref{conversed3}) in (\ref{pfe}),
we obtain  \be
    S = \frac{\Omega_{d-2}\,\rho_{0}^{d-2}}{4\,G_{d}}\,\cosh\alpha^{0}_{n}\,\cosh\alpha^{0}_{w}\( 1 + \frac {\(d-2\)^2} {2} \lambda'\)~,
 \ee which is eq. (\ref{full entropy}).

\end{document}